\documentclass[preprint,prd,nofootinbib,tightenlines,groupedaddress,amsmath,amssymb]{revtex4}
\usepackage{verbatim,graphics,graphicx,color,slashed}
\usepackage{ulem} 
\usepackage{hyperref}
\usepackage{changepage}

\def\lsim{\mathrel {\vcenter {\baselineskip 0pt \kern 0pt
    \hbox{$<$} \kern 0pt \hbox{$\sim$} }}}
\def\gsim{\mathrel {\vcenter {\baselineskip 0pt \kern 0pt
    \hbox{$>$} \kern 0pt \hbox{$\sim$} }}}
\def\slashchar#1{\setbox0=\hbox{$#1$}           
 \dimen0=\wd0                                 
  \setbox1=\hbox{/} \dimen1=\wd1               
\ifdim\dimen0>\dimen1                        
  \rlap{\hbox to \dimen0{\hfil/\hfil}}      
  #1                                        
  \else                                        
 \rlap{\hbox to \dimen1{\hfil$#1$\hfil}}   
   /                                         
  \fi}                                         %
\def\cpto{\mathrel {\vcenter {\baselineskip 0pt \kern 0pt
    \hbox{$CP$} \kern 0pt \hbox{$\longrightarrow$} }}}
\def\cptof{\mathrel {\vcenter {\baselineskip 0pt \kern 0pt
    \hbox{$~CP$} \kern 0pt \hbox{$\longleftrightarrow$} }}}

\begin{document}

\baselineskip=15pt

\preprint{}

\title{Some Predictions of Diquark Model for
 Hidden Charm \\Pentaquark Discovered at the LHCb }

\author{Guan-Nan Li${}^{1,3}$\footnote{lgn198741@126.com}}
\author{Xiao-Gang He$^{2,3,4}$\footnote{hexg@phys.ntu.edu.tw}}
\author{Min He$^{2}$\footnote{hemin\_sjtu@163.com}}

\affiliation{${}^1$ Department of Physics, Zhengzhou University, Henan, China}
\affiliation{${}^2$ INPAC, SKLPPC and Department of Physics and Astronomy,
  Shanghai Jiao Tong University, Shanghai, China}
\affiliation{${}^{3}$CTS, CASTS and
Department of Physics, National Taiwan University, Taipei, Taiwan}
\affiliation{${}^{4}$Physics Division, National Center for
Theoretical Sciences, Hsinchu, Taiwan}

\date{\today $\vphantom{\bigg|_{\bigg|}^|}$}

\date{\today}

\vskip 1cm
\begin{abstract}
The LHCb has discovered two new states with preferred $J^P$ quantum numbers $3/2^-$ and $5/2^+$ from $\Lambda_b$ decays. These new states can be interpreted as hidden charm pentaquarks.
It has been argued that the main features of these pentaquarks can be described by diquark model. The diquark model predicts that the $3/2^-$ and $5/2^+$ are in two separate octet multiplets of flavor $SU(3)$ and there is also an additional decuplet pentaquark multiplet. Finding the states in these multiplets can provide crucial evidence for this model.  The weak decays of b-baryon to a light meson and a pentaquark can have Cabibbo allowed and suppressed decay channels. We find that in the $SU(3)$ limit, for $U$-spin related decay modes the ratio of the decay rates of Cabibbo suppressed to Cabibbo allowed decay channels is given by $|V_{cd}|^2/|V_{cs}|^2$. There are also other testable relations for b-baryon weak decays into a pentaquark and a light pseudoscalar. These relations can be used as tests for the diquark model for pentaquark.
\end{abstract}
\pacs{PACS numbers: }

\maketitle


\noindent
{\bf Introduction}

The LHCb collaboration has recently discovered two new states\cite{Aaij:2015tga} which can be interpreted as two different pentaquarks ${\cal P}$ from $\Lambda_b\to {\cal P} + K$, followed by ${\cal P}\to J/\psi + p$. This has generated a lot of theoretical investigations\cite{rosner,Chen:2015loa,Chen:2015moa,Roca:2015dva,Feijoo:2015cca,Mironov:2015ica,Guo:2015umn,Maiani:2015vwa,He:2015cea,
Liu:2015fea,Lebed:2015tna,Mikhasenko:2015vca,xc,diquark1}. The quark contents can be identified as $\bar c c uud$. Although the states contain charm quarks, they are hidden charm pentaquarks because $c$ and $\bar c$ appear together and the net charm quantum number is zero. The best fit quantum numbers $J^P$ and their masses are
\begin{eqnarray}
J^P = 3/2^-\;\;\mbox{with a mass of 4380 MeV},\;\;\mbox{and}\;\; J^P = 5/2^+\;\;\mbox{with a mass of 4450 MeV}\;.
\end{eqnarray}
Experimentally quantum numbers $3/2^-$ and $5/2^+$ for these two states are not ruled out.

The existence of these states $P_c(4380)$ and $P_c(4450)$ need to be further confirmed as there may be some other effects which can minic similar effects\cite{Guo:2015umn,Liu:2015fea,Mikhasenko:2015vca}. If these states are genuine pentaquarks, one may ask whether they are molecular states of two hadrons or composite hadron systems\cite{rosner,Chen:2015loa,Chen:2015moa,Roca:2015dva,Mironov:2015ica,He:2015cea,xc}, or a tightly bound five quark system $[\bar c cqq'q'']$, or the quarks bound in other forms\cite{Maiani:2015vwa,Lebed:2015tna,diquark1}.
It is intriguing that the $\Sigma_c\bar D^*$ in the S-wave state has a mass very close to the $P_c(4450)$. Such a molecular state was actually studied just before the experimental discovery\cite{rosner}. With $\Sigma_c \bar D^*$ and $\Sigma_c^*\bar D^*$ bound sates one can obtain both $P_c(4380)$ and $P_c(4450)$ states\cite{Chen:2015loa}. These prompt the speculation that the pentaquarks might be molecular states. On the other hand, it has been argued that the two pentaquarks from the LHCb are five quark systems organized in $[q'q'']$ and $[c q]$ diquarks\cite{Maiani:2015vwa}, and the $\bar c$, the diquark model for pentaquark.  The diquark model also has supports from tetraquark studies\cite{Maiani:2015vwa}. At present, with limited data it is not possible to distinguish whether the pentaquarks are molecular states or more tightly bound quark states or even mixture of these states. The diquark model has simple structure to analysis. We also find it very predictive. Therefore  in this work we choose to study some properties of diquark model for the pentaquarks discovered at the LHCb.
\\

\noindent
{\bf Diquark Model for Pentaquarks}

In the  diquark model the two pentaquarks from the LHCb are five quark systems organized in $[q'q'']$ and $[c q]$ diquarks\cite{Maiani:2015vwa}, and the $\bar c$. More explicitly indicated as the following
\begin{eqnarray}
{\cal P} = \epsilon^{\alpha \beta\gamma}\{\bar c_\alpha [cq]_{\beta,\;s=0,1} [q'q'']_{\gamma,\;s=0,1}, L\}\,
\end{eqnarray}
where the greek letters are color indices and $s$ indicates the spin.

Under flavor $SU(3)$ symmetry, the diquark $[q'q'']$ transforms as $\bar 3$ and $6$ and the diquark $[cq]$ transforms as a $3$. Therefore the pentaquarks can have $3\times \bar 3 = 1 + 8$ and $3\times 6 = 8 + 10$ multiplets.
We indicate the pentaquarks with $\bar 3$ and $6$ by ${\cal P}_A$ and ${\cal P}_S$, respectively.
Assuming that the two pentaquarks have $J^P = 3/2^-$ and $J^P = 5/2^+$,
these two fields are component fields in octet multiplets with the following spin diquark combinations
fit the picture well\cite{Maiani:2015vwa}
\begin{eqnarray}
&&{\cal P}_S(3/2^-) = \epsilon^{\alpha \beta\gamma}\{\bar c_\alpha [cq]_{\beta,\;s=1} [q'q'']_{\gamma,\;s=1}, L=0\}\,\nonumber\\
&&{\cal P}_A(5/2^+) = \epsilon^{\alpha \beta\gamma}\{\bar c_\alpha [cq]_{\beta,\;s=1} [q'q'']_{\gamma,\;s=0}, L=1\}\,
\end{eqnarray}

We denote the octet pentaquark component fields as
\begin{eqnarray}
({\cal P}_{i}^{j}(J^P)) = \left ( \begin{array}{ccc}
{\Sigma^{0}_8\over \sqrt{2}}+{\Lambda^{0}_8 \over \sqrt{6}}& \Sigma^{+}_8& p_8\\
\Sigma^{-}_8& -{\Sigma^{0}_8\over \sqrt{2}}+{\Lambda^{0}_8 \over \sqrt{6}}&n_8\\
\Xi^{-}_8&  \Xi^{0}_8& -{2\Lambda^{0}_8\over \sqrt{6}}
\end{array}
\right )\;.
\end{eqnarray}

For $J^P = 3/2^-$, there should also be a decuplet ${\cal P}_{ijk}$ (totally symmetric in sub-indices) multiplet. The component pentaquark fields are
\begin{eqnarray}
&&{\cal P}_{111} = \Delta^{++}_{10}\;,\;\;\;\;{\cal P}_{112} = {1\over \sqrt{3}}\Delta^{+}_{10}\;,\;\;\;{\cal P}_{122} = {1\over \sqrt{3}}\Delta^{0}_{10}\;,\;\;\;\;{\cal P}_{222} = \Delta^{-}_{10}\;,\nonumber\\
&&{\cal P}_{113} = {1\over \sqrt{3}}\Sigma^{+}_{10}\;,\;\;\;\;{\cal P}_{123} = {1\over \sqrt{6}}\Sigma^{0}_{10}\;,\;\;\;\;{\cal P}_{223} = {1\over \sqrt{3}}\Sigma^{-}_{10}\;,\nonumber\\
&&{\cal P}_{133} = {1\over \sqrt{3}}\Xi^{0}_{10}\;,\;\;\;\;{\cal P}_{233} = {1\over \sqrt{3}}\Xi^{-}_{10}\;,\nonumber\\
&&{\cal P}_{333} = \Omega^{-}_{10}\;.
\end{eqnarray}

The two observed pentaquarks are identified as $p_8(3/2^-)$ and $p_{8}(5/2^+)$, respectively. It is clear that  there are other members of pentaquarks. Similar to the decay channels $p_8(3/2^-, 5/2^+) \to J/\psi + p$, the pentaquarks in the octet and decuplet will be able to decay into a $J/\psi$ plus baryons in the low-lying octet and decuplet, respectively. For $J^P=5/2^+$, there should be a companion singlet pentaquark $\cal S$. However, there is no low-lying baryon singlet, ${\cal S}$ will not be able to decay into a $J/\psi$ plus an ordinary low-lying baryon, but may be an ordinary baryon and multi-mesons forming a $SU(3)$ singlet. The masses of the pentaquarks are degenerate in the $SU(3)$ limit. But may be different due to s-quark mass being much larger than u- and d-quarks. Estimate of mass differences for some of the pentaquarks in the diquark model has been carried out in Ref.\cite{diquark1}.
Discoverying these additional pentaquarks identified above is one of the way to verify the diquark model for pentaquark which can in principle be carried at the LHCb.
\\

\noindent
{\bf Pentaquark Weak Decays\;\;}

We now discuss weak decay modes of b-baryon into an octet or a decuplet pentaquark and a light pseudoscalar octet meson. The leading effective Hamiltonian inducing
$\cal B \to \cal M + \cal P$ decays in the SM has both parity  conserving and violating parts given by
\begin{eqnarray}
 H_{eff}(q) = {4 G_{F} \over \sqrt{2}} [V_{cb}V^{*}_{cq} (c_1 O_1 +c_2 O_2),
\end{eqnarray}
where $q$ can be $d$ or $s$. $V_{ij}$ is the CKM matrix element. The coefficients
$c_{1,2}$ are the Wilson Coefficients (WC) which WC have been studied by several groups and can be
found in Ref.~\cite{heff}. The operators $O_i$ are
 given by
\begin{eqnarray}
O_1=(\bar q_i c_j)_{V-A}(\bar c_i b_j)_{V-A}\;,\;\;\;\;
O_2=(\bar q c)_{V-A}(\bar c b)_{V-A}\;,
\end{eqnarray}
where $(\bar a b)_{V\pm A} = \bar a \gamma_\mu (1\pm \gamma_5) b$. In the above, we have neglected contributions from penguin diagrams since they are significantly smaller than the tree contributions given above.

The operators $O_{1,2}$ transfer under the  flavor $SU(3)$ as a $\bar 3$. We indicate it as $ H(\overline{3})$.
The non-zero entries of the matrices $H(\overline{3})$ are given as the following,
\begin{eqnarray}
&&H(\overline{3})^{2}=1\;, \;\;\;\;\mbox{for $\Delta S=0,\;\;\;q=d$}\;,\nonumber\\
&&H(\overline{3})^{3}=1\;,\;\;\;\;\mbox{for $\Delta S=-1, q=s$}\;.
\end{eqnarray}

For $q=s$, the decay amplitude is proportional to $V_{cs}$, which we refer to as Cabibbo allowed interaction.  For $q=d$,
the decay amplitude is proportional to $V_{cd}$, which is Cabibbo suppressed interaction.

The low-lying ${1\over 2}^+$ $\cal B$ b-baryons are made up of a $b$ quark and two light quarks. Here the light quark $q$ is one of the $u$, $d$ or $s$ quarks.
Under the flavor $SU(3)$ symmetry, the $b$ quark is a singlet and the light quark $q$ is a member in the fundamental representation $3$.  The b-baryons then have representations under flavor $SU(3)$ as $1\times 3 \times 3 \time 3 = \bar 3 + 6$, that is,  the b-baryons contain an anti-triplet and a sextet in the  $SU(3)$ flavor space\cite{b-baryon}.  The anti-triplet $\cal B$ and the sextet ${\cal C}$ b-baryons will be indicated by
 \begin{eqnarray}
 ({\cal B}_{ij})=\left ( \begin{array}{ccc}0&\Lambda^{0}_{b}&\Xi^{0}_{b}\\
 -\Lambda^{0}_{b}&0&\Xi^{-}_{b}\\
 -\Xi^{0}_{b}&-\Xi^{-}_{b}&0
 \end{array}
\right )\;,\;\;({\cal C}_{ij})=\left ( \begin{array}{ccc}
\Sigma^{+}_{b}& {\Sigma^{0}_{b}\over \sqrt{2}} &{\Xi'^{0}_{b}\over \sqrt{2}}\\
{\Sigma^{0}_{b}\over \sqrt{2}} & \Sigma^{-}_{b} &{\Xi'^{-}_{b}\over \sqrt{2}}\\
{\Xi'^{0}_{b}\over \sqrt{2}} &  {\Xi'^{-}_{b}\over \sqrt{2}} &\Omega^{-}_{b}
\end{array}
\right )\;.
 \end{eqnarray}

The pseudoscalar octet mesons will be indicated  by
$\cal M$. They are
\begin{eqnarray}
({\cal M}_{i}^{j}) = \left ( \begin{array}{ccc}
{\pi^{0}\over \sqrt{2}}+{\eta_{8} \over \sqrt{6}}& \pi^{+}& K^{+}\\
\pi^{-}& -{\pi^{0}\over \sqrt{2}}+{\eta_{8} \over \sqrt{6}}&{K}^{0}\\
K^{-}&  \bar K^{0} & -{2\eta_{8} \over \sqrt{6}}
\end{array}
\right )\;.
\end{eqnarray}

At the hadron level, the decay amplitude can be generically written as
\begin{eqnarray}
{\cal A} = \langle {\cal P} {\cal M}\vert H_{eff}(q)\vert {\cal {B}}\;\; \mbox{or}\;\; {\cal C}\rangle =  V_{cb}V^*_{cq} T(q).
\end{eqnarray}

To obtain the $SU(3)$ invariant decay amplitude for a b-baryon, one first uses the Hamiltonian to annihilate the b-quark in ${\cal B}$ (or ${\cal C}$)  and then contract $SU(3)$ indices in an appropriate way with final states $\cal P$ and $\cal M$.  Taking the anti-triplet tree amplitude $T_t(q)$ and sextet tree amplitude $T_s(q)$ as examples, following the procedures for b-baryon charmless two-body decays in Ref.\cite{he-b-baryon}, we have

\begin{eqnarray}
T_{t8}(q)&=&a_t(\overline{3})\langle {\cal P}^{k}_{l}{\cal M}^{l}_{k}\vert H(\overline{3})^{i} \vert {\cal B}_{i'i''}\rangle \epsilon ^{ii'i''}
+b_t(\overline{3})\langle {\cal P}^{k}_{j} {\cal M}^{i}_{k} \vert H(\overline{3})^{j}\vert {\cal B}_{i'i''}\rangle \epsilon ^{ii'i''}\nonumber\\
&+&c_t(\overline{3})\langle {\cal P}^{i}_{k}{\cal M}^{k}_{j}\vert H(\overline{3})^{j} \vert {\cal B}_{i'i''}\rangle \epsilon ^{ii'i''}\nonumber\\
&+&d_t(\overline{3})\langle {\cal P}^{i'}_{j'}{\cal M}^{i}_{j} \vert H(\overline{3})^{i''}  \vert {\cal B}_{jj'}\rangle \epsilon_{ii'i''}
+e_t(\overline{3})\langle {\cal P}^{i'}_{j'}{\cal M}^{i}_{j}\vert H(\overline{3})^{j}\vert {\cal B}_{i''j'}\rangle\epsilon_{ii'i''}\nonumber\\
&+&f_t(\overline{3}) \langle {\cal P}^{i'}_{j}{\cal M}^{i}_{j'}\vert H(\overline{3})^{j}\vert {\cal B}_{i''j'}\rangle\epsilon_{ii'i''}\;.
\end{eqnarray}
and
\begin{eqnarray}
T_{s8}(q)&=&d_s(\overline{3})\langle {\cal P}^{i'}_{j'} {\cal M}^{i}_{j}\vert H(\overline{3})^{i''}  \vert {\cal C}_{jj'}\rangle \epsilon_{ii'i''}
+e_s(\overline{3})\langle {\cal P}^{i'}_{j'}{\cal M}^{i}_{j}\vert H(\overline{3})^{j}\vert {\cal C}_{i''j'}\rangle\epsilon_{ii'i''}\nonumber\\
&+&f_s(\overline{3}) \langle {\cal P}^{i'}_{j}{\cal M}^{i}_{j'}\vert H(\overline{3})^{j}\vert {\cal C}_{i''j'}\rangle\epsilon_{ii'i''}\;.
\end{eqnarray}

The $SU(3)$ invariant decay amplitude involve decuplet pentaquarks can be written as
\begin{eqnarray}
T_{t10}(q)&=&a_{t10}\langle {\cal P}_{kjl}{\cal M}^{k}_{i}\vert H(\overline{3})^{l} \vert {\cal B}_{ij}\rangle\;,\nonumber\\
T_{s10}(q)&=&a_{s10}\langle {\cal P}_{kjl}{\cal M}^{k}_{i}\vert H(\overline{3})^{l} \vert {\cal C}_{ij}\rangle + b_{s10}\langle {\cal P}_{kji}{\cal M}^{k}_{l}\vert H(\overline{3})^{l} \vert {\cal C}_{ij}\rangle\;.\label{decuplet}
\end{eqnarray}

For ${\cal B}$ baryons decay into $\cal P$, there are six possible terms as far as $SU(3)$ properties are concerned.
The dominant contributions are from terms with coefficients $c_t(\overline {3})$, $d_t(\overline {3})$, $e_t(\overline{3})$ and $f_t(\overline {3})$. In figure 1, we show diagrams correspond to these terms. In these figures, the quarks $q_i$, $q_{i'}$ and $q_{i''}$ are contracted by the totally antisymmetric tensor $\epsilon^{ii'i''}$ or $\epsilon_{ii'i''}$. This property plays an important role in determining whether $\bar 3$ or sextet diquarks are allowed to form. The two terms
with coefficients $a_t(\overline{3})$ and $b_t(\overline {3})$ are allowed, but it turns out that for these two terms the
$\cal P$ state is actually a higher order fork state with the same $SU(3)$ quantum numbers but with more quark contents as can be seen in figure 2. Therefore, one expects that the coefficients  $a_t(\overline{3})$ and $b_t(\overline {3})$ to be smaller than
$c_t(\overline {3})$, $d_t(\overline {3})$, $e_t(\overline{3})$ and $f_t(\overline {3})$. We will include all in our later discussions. For the properties we emphasis will not be affected whether to include $a_t(\overline{3})$ and $b_t(\overline {3})$ or not.

For ${\cal C}$ baryons decay into $\cal P$, only three possible terms, terms with coefficients, $d_s(\overline {3})$, $e_s(\overline{3})$ and $f_s(\overline {3})$. The corresponding diagrams are shown in figure 1.b, 1.c, and 1.d. Because the two light quarks in $\cal C$ are symmetric, similar terms to those with coefficients $a_t(\overline {3})$, $b_t(\overline {3})$, $c_t(\overline{3})$ for $\cal B$ decays are identically zero.

For $\cal B$ and $\cal C$ baryons decay into decuplet, the final light quarks need to be in totally symmetric state. There are less possibilities. In figure 3 we show the corresponding diagrams for the allowed terms.

\begin{figure}[!h]
\begin{centering}
\begin{tabular}{cc}
\includegraphics[width=0.45\textwidth]{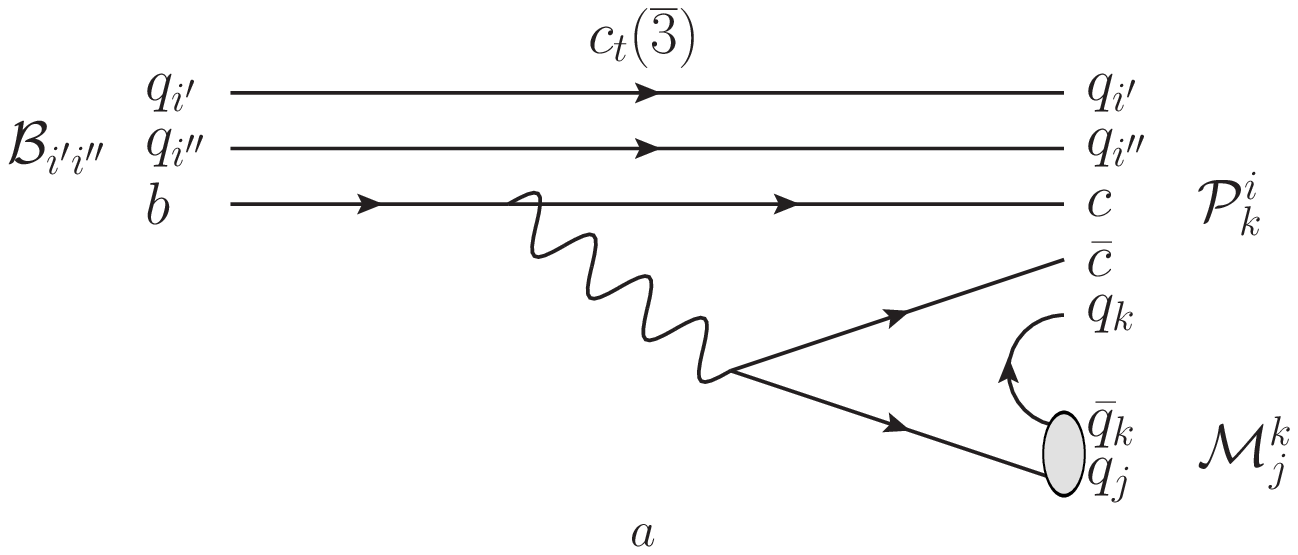}
\hspace*{1em}
\includegraphics[width=0.45\textwidth]{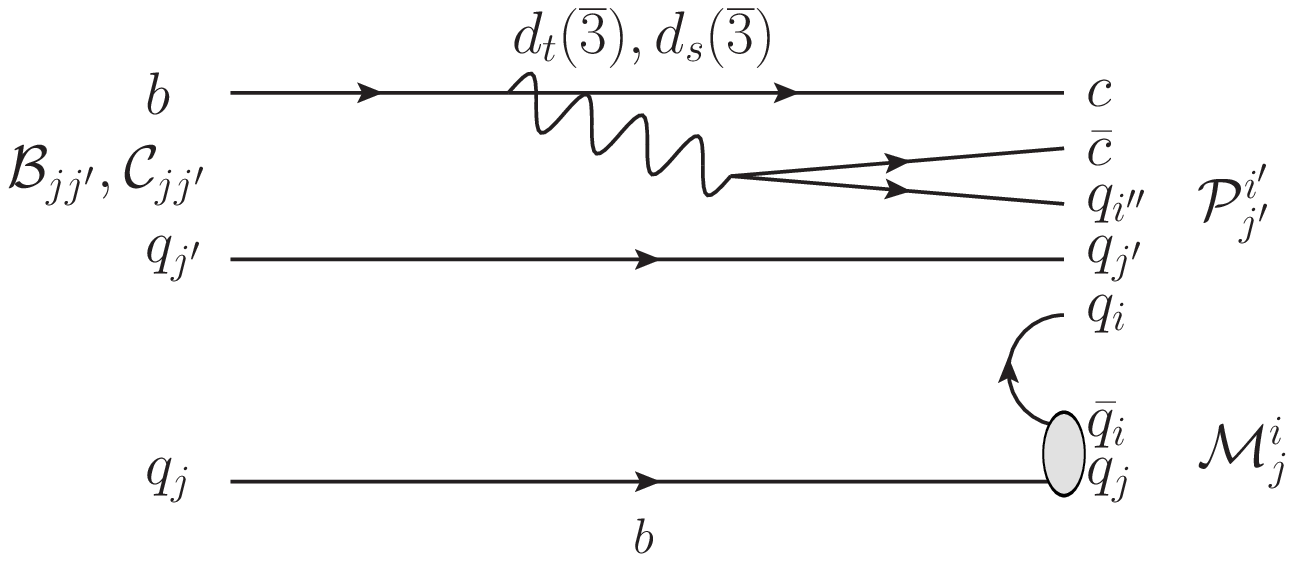}
\\
\includegraphics[width=0.45\textwidth]{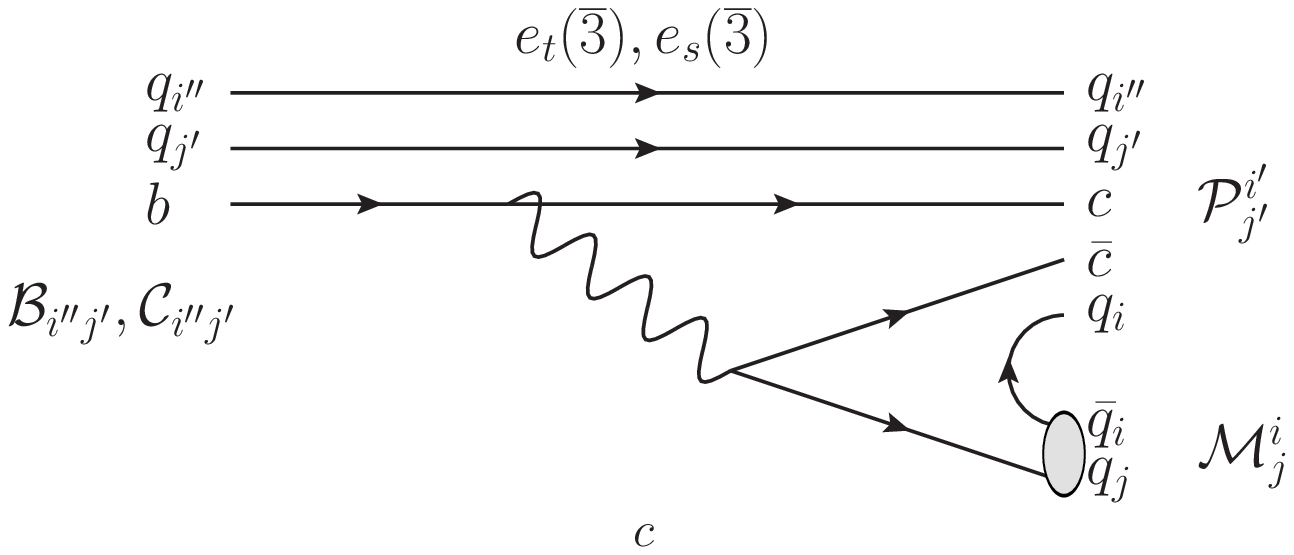}
\hspace*{1em}
\includegraphics[width=0.45\textwidth]{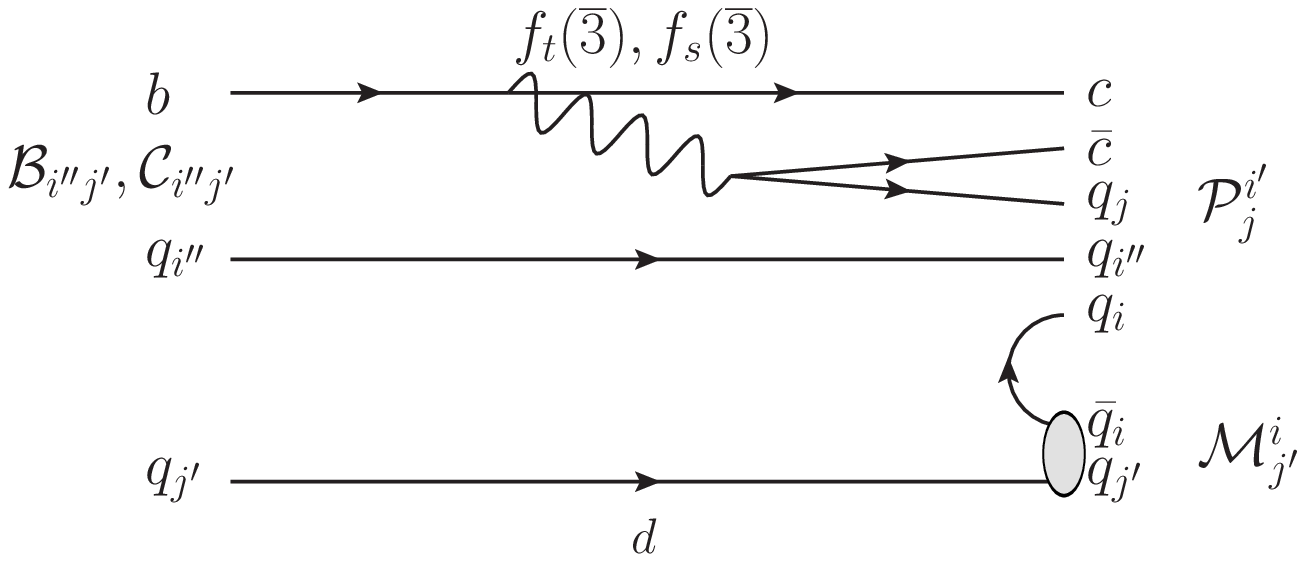}
\end{tabular}
\end{centering}
\caption{
Figure 1.a corresponds to the term with coefficient $c_t(\overline{3})$. $\overline{3}$ diquarks can be formed by $(q_{i'}q_{i''})$ pair only. The pentaquark is formed by  $c\bar c q_k q_{i'}q_{i''}$.
Figure 1.b corresponds to the term with coefficients $d_t(\overline{3})$ and $d_s(\overline {3})$. Possible $\overline{3}$ diquarks can be formed by $(q_{i} q_{i''})$, $(q_iq_{j'})$ and $(q_{i''}q_{j'})$ pairs. Possible sextet diquark can be formed by $(q_{i''} q_{j'})$ and $(q_i q_{j'})$. The pentquark is formed by $c\bar cq_i q_{i''}q_{j'}$. Figure 1.c corresponds to the term with coefficients $e_t(\overline{3})$ and $e_s(\overline {3})$. Possible $\overline{3}$ diquarks can be formed by $(q_i q_{i''})$, $(q_{i}q_{j'})$ and $(q_{i''}q_{j'})$ pairs. Possible sextet diquark can be formed by $(q_i q_{j'})$ and $(q_{i''} q_{j'})$. The pentquark is formed by $c\bar cq_i q_{i''}q_{j'}$. Figure 1.d corresponds to the term with coefficients $f_t(\overline{3})$ and $f_s(\overline {3})$. Possible $\overline{3}$ diquarks can be formed by $(q_i q_{i''})$, $(q_{i}q_{j})$ and $(q_{i''}q_{j})$ pairs. Possible sextet diquark can be formed by $(q_i q_{j})$ and $(q_{i''} q_{j})$. The pentquark is formed by $c\bar cq_i q_{i''}q_{j}$.
}
\end{figure}


\begin{figure}[!h]
\begin{centering}
\begin{tabular}{cc}
\includegraphics[width=0.45\textwidth]{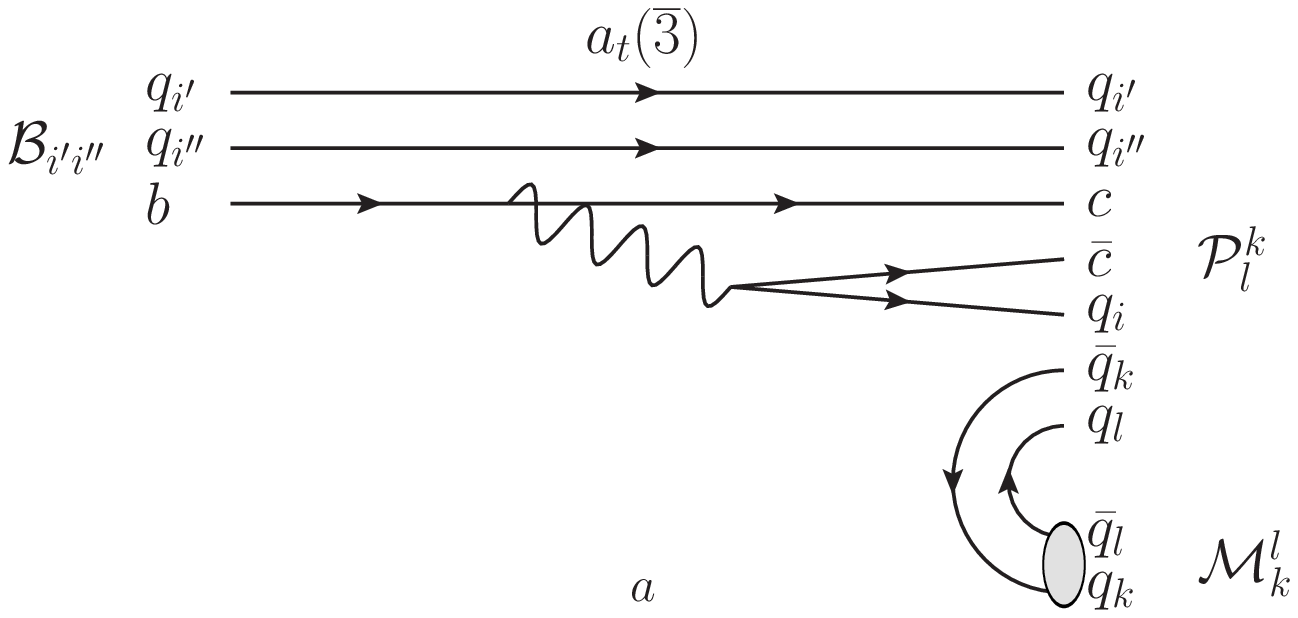}
\hspace*{1em}
\includegraphics[width=0.45\textwidth]{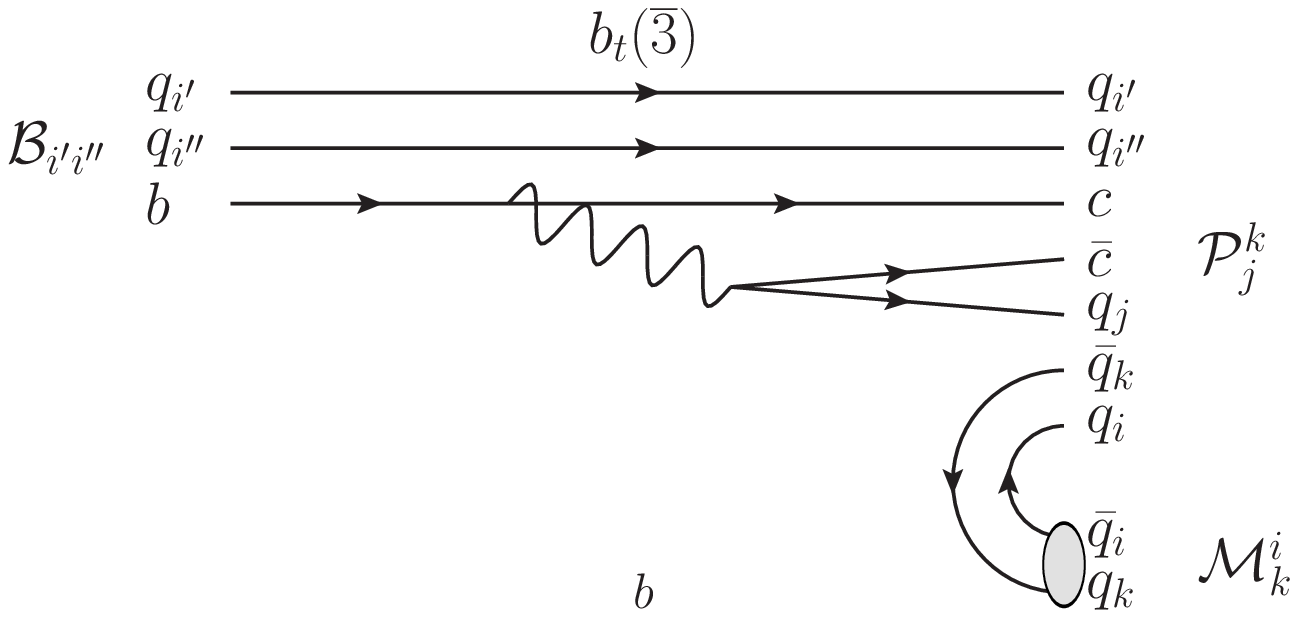}
\end{tabular}
\end{centering}
\caption{
In the above two figures, the quarks $q_{i'}q_{i''}$ are in $\bar 3$ flavor state.
Figure 2.a corresponds to the term with coefficient $a_t(\overline{3})$. Possible $\overline{3}$ diquarks can be formed by $(q_i q_{i'})$, $(q_iq_{i''})$ and $(q_{i'}q_{i''})$ pairs. The ${\cal{P}}^k_l$ can be viewed as a singlet pentaquark formed by $c \bar c q_iq_{i'}q_{i''}$ bounded with an octet $(\bar q_k q_l)$ state. Figure 2.b corresponds to the term with coefficient $b_t(\overline{3})$. Possible $\overline{3}$ diquarks can be formed by $(q_i q_{i'})$, $(q_iq_{i''})$ and $(q_{i'}q_{i''})$ pairs. The ${\cal{P}}^k_j$ can be viewed as a pentaquark formed by $c \bar c q_{i'}q_{i''}q_i $ bounded with an octet $(\bar q_k q_j)$ state.
In Figures 2.a and 2.b the $\cal P$ states having the same quantum number as the LHCb observed pentaqaurk are higher order fork states. The corresponding coefficients $a_t(\overline{3})$ and $b_t(\overline{3})$ are expected to be smaller than $c_t(\overline {3})$.
}
\end{figure}

\begin{figure}[h!]
\begin{centering}
\begin{tabular}{c}
\includegraphics[width=0.45\textwidth]{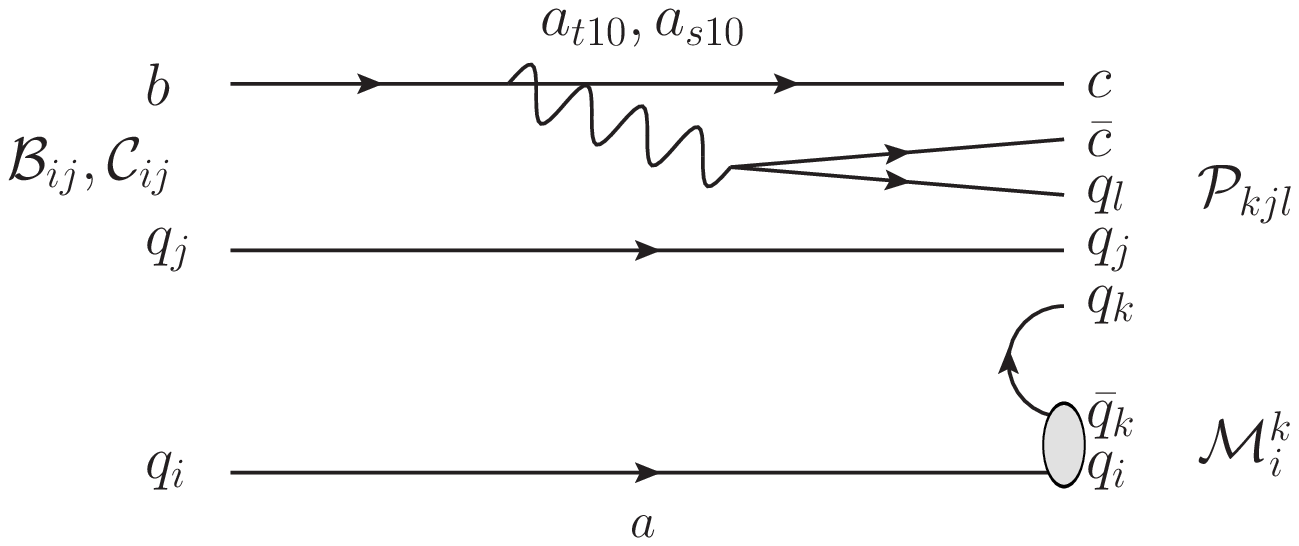}
\hspace*{2em}
\includegraphics[width=0.45\textwidth]{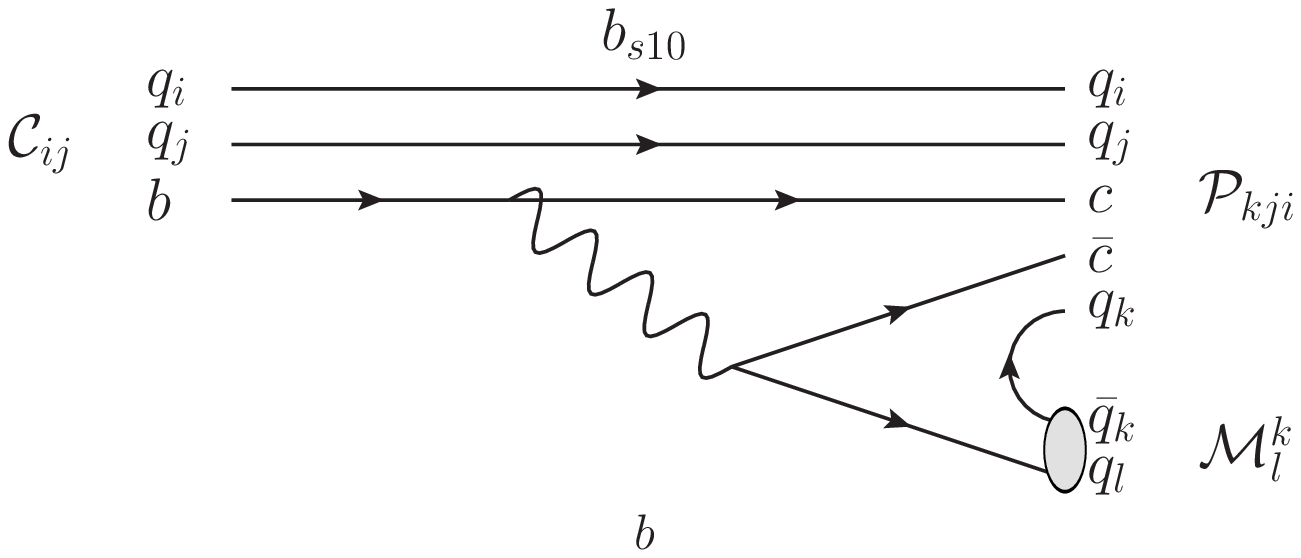}
\end{tabular}
\caption{
Figure 3.a corresponds to the term with coefficients $a_{t10}$ and $a_{s10}$. Possible sextet diquarks can be formed by $(q_j q_k)$, $(q_jq_l)$ and $(q_kq_l)$ pairs. Figure 3.b corresponds to the term with coefficient $b_{s10}$. Possible sextet diquarks can be formed by $(q_iq_j)$, $(q_iq_k)$ and $(q_jq_k)$ pairs.
}
\end{centering}
\end{figure}


In the above, we have suppressed the Lorentz indices and spinor forms, but concentrated in $SU(3)$ flavor indices. The results apply to $3/2^-$ and also $5/2^-$ multiplets.
Expanding the above amplitudes, one can obtain the individual decay amplitude. The full expansions are given in the appendices.  From these results, one can read off many properties concerning weak decays of b-baryons to a pentaquark and a light pseudoscalar. We present some of the interesting properties in the following.
\\

\noindent
{\bf Discussions and conclusions}

We begin with discussing b-baryons decay into octet pentaquarks. The two pentaquarks discovery at the LHCb belong to this category. Without a detailed dynamic model, one can not calculate the size of the various $SU(3)$ amplitudes, but the flavor $SU(3)$ symmetry can relate to different decay modes and can be tested experimentally. Since the $J^P = 3/2^-$ and $J^P = 5/2^+$ belong to different octet, there are no relations among the decay amplitudes related to these two pentaquarks. But within each multiplet with the same $J^P$, there are relations which can be tested experimentally.

We find the $U$-spin related amplitudes ($\Delta S=0$ and $\Delta S=-1$) for anti-triplet satisfy the following relations
\begin{eqnarray}
&& T_t(\Xi^{-}_{b} \to K^{-} n_8)= T_t(\Xi^{-}_{b} \to \pi^{-} \Xi^{0}_8)\;, \;\;\;\;\;\;\;\;
 T_t(\Xi^{0}_{b} \to \bar{K}^{0} n_8)=-T_t(\Lambda^{0}_{b} \to K^{0} \Xi^{0}_8)\;, \nonumber\\
&& T_t(\Xi^{-}_{b} \to K^{0} \Xi^{-}_8)=T_t(\Xi^{-}_{b} \to \bar{K}^{0} \Sigma^{-}_8)\;, \;\;\;\;\;\;\;
 T_t(\Xi^{0}_{b} \to K^{0} \Xi^{0}_8)=-T_t(\Lambda^{0}_{b} \to \bar{K}^{0} n_8)\;, \nonumber\\
&& T_t(\Xi^{0}_{b} \to \pi^{-} \Sigma^{+}_8)=-T_t(\Lambda^{0}_{b} \to K^{-} p_8)\;, \;\;\;\;\;\;
 T_t(\Lambda^{0}_{b} \to \pi^{-} p_8)=-T_t(\Xi^{0}_{b}\to K^{-} \Sigma^{+}_8)\;,\nonumber\\
&& T_t(\Xi^{0}_{b} \to \pi^{+} \Sigma^{-}_8)=-T_t(\Lambda^{0}_{b} \to K^{+} \Xi^{-}_8)\;,\;\;\;\;\;
  T_t(\Lambda^{0}_{b} \to K^{+} \Sigma^{-}_8)=-T_t(\Xi^{0}_{b} \to \pi^{+} \Xi^{-}_8)\;,\nonumber\\
&& T_t(\Xi^{0}_{b} \to K^{-} p_8)=-T(\Lambda^{0}_{b} \to \pi^{-} \Sigma^{+}_8)\;,\;\;\;\;\;\;\;
  T_t(\Xi^{0}_{b} \to K^{+} \Xi^{-}_8)=-T_t(\Lambda^0_{b} \to \pi^{+} \Sigma^{-}_8)\;. \label{r1}
\end{eqnarray}
While the $U$-spin related amplitudes for sextet satisfy
\begin{eqnarray}
&& T_s(\Sigma^{+}_{b} \to n_8 \pi^{+})=-T_s(\Sigma^{+}_{b} \to \Xi^{0}_8 K^{+})\;,\;\;\;\;
T_s(\Sigma^{+}_{b} \to \Sigma^{+}_8 K^{0})=-T_s(\Sigma^{+}_{b} \to p_8 \bar{K}^{0})\;,\nonumber\\
&& T_s(\Sigma^{-}_{b} \to n_8 \pi^{-})=-T_s(\Omega^{-}_{b} \to \Xi^{0}_8 K^{-})\;,\;\;\;\;
 T_s(\Sigma^{-}_{b} \to \Sigma^{-}_8 K^{0})=-T_s(\Omega^{-}_{b} \to \Xi^{-}_8 \bar{K}^{0})\;,\nonumber\\
&& T_s(\Omega^{-}_{b} \to \Xi^{0}_8 \pi^{-})=-T_s(\Sigma^{-}_{b} \to n_8 K^{-})\;,\;\;\;\;
 T_s(\Omega^{-}_{b} \to \Sigma^{-}_8 \bar{K}^{0})=-T_s(\Sigma^{-}_{b} \to \Xi^{-}_8 K^{0})\;,\nonumber\\
&& T_s(\Sigma^{0}_{b} \to \Sigma^{-}_8 K^{+})=-T_s(\Xi'^{0}_{b} \to \Xi^{-}_8 \pi^{+})\;,\;\;\;
 T_s(\Sigma^{0}_{b} \to p_8 \pi^{-})=-T_s(\Xi'^{0}_{b} \to \Sigma^{+}_8 K^{-})\;,\nonumber\\
&&  T_s(\Xi'^{0}_{b} \to \Xi^{-}_8 K^{+})=-T_s(\Sigma^{0}_{b} \to \Sigma^{-}_8 \pi^{+})\;,\;\;\;
 T_s(\Xi'^{0}_{b} \to \Sigma^{-}_8 \pi^{+})=-T_s(\Sigma^{0}_{b} \to \Xi^{-}_8 K^{+})\;,\nonumber\\
&& T_s(\Xi'^{0}_{b} \to p_8 K^{-})=-T_s(\Sigma^{0}_{b} \to \Sigma^{+}_8 \pi^{-})\;,\;\;\;\;
  T_s(\Xi'^{0}_{b} \to \Sigma^{+}_8 \pi^{-})=-T_s(\Sigma^{0}_{b} \to p_8 K^{-})\;,\nonumber\\
&& T_s(\Xi'^{0}_{b} \to \Xi^{0}_8 K^{0})=-T_s(\Sigma^{0}_{b} \to n_8 \bar{K}^{0})\;,\;\;\;\;\;
 T_s(\Xi'^{0}_{b} \to n_8 \bar{K}^{0})=-T_s(\Sigma^{0}_{b} \to \Xi^{0}_8 K^{0})\;,\nonumber\\
&& T_s(\Xi'^{-}_{b} \to n_8 K^{-})=-T_s(\Xi'^{-}_{b} \to \Xi^{0}_8 \pi^{-})\;,\;\;\;
 T_s(\Xi'^{-}_{b} \to \Xi^{-}_8 K^{0})=-T_s (\Xi'^{-}_{b} \to \Sigma^{-}_8 \bar{K}^{0}). \label{r2}
\end{eqnarray}

The above relations can be directly read off from the diagrams shown in figures 1 and 2 when specific quarks are assigned for each decay modes.
For illustrations we take the pairs  a). $\Xi^0_b \to K^+ \Xi^-_8$ and $\Lambda_b^0 \to \pi^+ \Sigma^-_8$ for $\cal B$ decay, and b). $\Omega^-_b \to \pi^- \Xi^0_8$ and $\Sigma^-_b\to K^- n_8$ for $\cal C$ decay  as examples to provide some details.
From tables in Appendix A, we find that for the pair in a), there are contributions from $a_t(\overline{3})$ and $d_t(\overline{3})$ terms. We focus on the diagram corresponds to $d_t(\overline {3})$.
For the pair in b), there is only $e_s(\overline {3})$ contribution. Specifying quark contents, we
have
\begin{eqnarray}
a) &&\mbox{For}\;\; \Xi^0_b \to K^+ \Xi^-_8\;,\;\;(q_{j'},\;q_j,\;q_i,\;q_{i''}) = ( s,\;u,\;s,\;d)\;,\nonumber\\
&&\mbox{For}\;\; \Lambda_b^0 \to \pi^+ \Sigma^-_8\;,\;\;\;(q_{j'},\;q_j,\;q_i,\;q_{i''}) = ( d,\;u,\;d,\;s)\;. \nonumber\\
b) &&\mbox{For}\;\; \Omega^-_b \to \pi^- \Xi^0_8\;,\;\;\;(q_{j'},\;q_j,\;q_i,\;q_{i''}) = ( s,\;d,\;u,\;s)\;,\nonumber\\
&&\mbox{For}\;\; \Sigma^-_b\to K^- n_8\;,\;\;(q_{j'},\;q_j,\;q_i,\;q_{i''}) = ( d,\;s,\;u,\;d)\;.
\end{eqnarray}

Note that for each pairs one just needs to exchange all s quarks with all d quarks to go from one to another within a given pair. If U spin is a good symmetry, the amplitudes defined in eq.15 and eq.16 are therefore equal in magnitude. The relative minus sign for the amplitude in each pair in eq.15 and eq.16 comes from the fact that each diagram is contracted by $\epsilon^{ii'i''}$ and $\epsilon_{ii'i''}$ for $\cal B$ and $\cal C$ decays, respectively. For the pair in a), $(i=3, i'=1, i''=2)$ and $(i=2, i'=1, i''=3)$ for $\Xi_b^0$ and $\Lambda^0_b$ decays, and for the pair in b), $(i=1, i'=2, i''=3)$ and $(i=1, i'=3, i''=2)$ for $\Omega_b^-$ and $\Sigma^-_b$ decays, respectively. These specific values for $i$, $i'$ and $i''$ explain the relative minus sign in the relations.

The above relations also hold even one include small penguin contributions\cite{he-b-baryon}.
These relations apply to both octets with $J^P = 3/2^-$ and $J^P = 5/2^+$.
Due to mixing between $\eta_8$ and $\eta_1$, the decay modes with $\eta_8$ in the final sates is not as clean as those with $\pi$ and $K$ in the final state to study. We have not listed processes involve $\eta_8$ above.

The above relations lead to the following relations for each pairs above,
\begin{eqnarray}
A({\cal B} \to M {\cal P}, \Delta S = 0) = V_{cb}V_{cd}^* T\;,\;\;A({\cal B} \to M {\cal P}, \Delta S = -1) = V_{cb}V_{cs}^* T\;,
\end{eqnarray}
and
\begin{eqnarray}
{\Gamma({\cal B} \to M {\cal P}, \Delta S = 0)\over \Gamma({\cal B} \to M {\cal P}, \Delta S = -1)} = {|V_{cd}|^2\over |V_{cs}|^2} \approx 4.5\%\;.
\end{eqnarray}
When more data become available, with more pentaquarks discovered, the above relations can be tested. To study the Cabbibo suppressed decays, 20 times more data are needed.

For the b-baryons decay into decuplet pentaquark, let us focus on b-baryons which undergo visible weak decays, namely $\Lambda_b^0$, $\Xi^0_b$, $\Xi_b^-$, and $\Omega^-_b$ decays. The full lists are given in the appendix.
The b-baryons  $\Lambda_b^0$, $\Xi^0_b$, $\Xi_b^-$ belong to the anti-triplet ${\cal B}$. Expanding the first equation in eq.(\ref{decuplet}), we have for $\Delta S = 0$ amplitudes
\begin{eqnarray}
\Lambda_b^0: &&a_{t10} (\pi^+\Delta^-_{10} + {1\over \sqrt{3} }K^+ \Sigma^-_{10}+ {\sqrt{2}\over \sqrt{3}}\pi^0 \Delta^0_{10}  -{1\over \sqrt{3}}\pi^- \Delta_{10}^+ - {1\over \sqrt{6}}  K^0 \Sigma_{10}^0)\nonumber\\
\Xi^0_b:&&a_{t10}({1\over 2 \sqrt{3} }\pi^0\Sigma^0_{10} + {1\over 2} \eta_8 \Sigma^0_{10}
+{1\over \sqrt{3}} \pi^+ \Sigma^-_{10} + {1\over \sqrt{3}} K^+ \Xi^-_{10} - {1\over \sqrt{3}} K^- \Delta^+_{10} - {1\over \sqrt{3}} \bar K^0 \Delta^0_{10})\;,\nonumber\\
\Xi^-_{b}:&&a_{t10}({1\over \sqrt{6}}\pi^- \Sigma^0_{10} - {1\over \sqrt{6}} \pi^0 \Sigma^-_{10} +{1\over \sqrt{2}}\eta_8 \Sigma^-_{10} + {1\over \sqrt{3}} K^0 \Xi^-_{10} - {1\over \sqrt{3}} K^- \Delta^0_{10} -\bar K^0 \Delta^-_{10})\;,
\end{eqnarray}
and for $\Delta S= - 1$ amplitudes, we have
\begin{eqnarray}
\Lambda^0_b: &&a_{t10}({1\over \sqrt{3}}\pi^0\Sigma^0_{10}  + {1\over \sqrt{3}}\pi^+ \Sigma^-_{10} +
{1\over \sqrt{3}} K^+ \Xi^-_{10} -{1\over \sqrt{3}} \pi^- \Sigma_{10}^+ - {1\over \sqrt{3}}  K^0 \Xi_{10}^0)\;,\nonumber\\
\Xi^0_{b}:&&a_{t10}({1\over \sqrt{6}} \pi^0\Xi^0_{10} + {1\over \sqrt{2}} \eta_8 \Xi^0_{10} + {1\over \sqrt{3}} \pi^+ \Xi^-_{10} + K^+ \Omega^-_{10} - {1\over \sqrt{3}} K^- \Sigma^+_{10} - {1\over \sqrt{6}}\bar K^0 \Sigma^0_{10})\;,\nonumber\\
\Xi^-_{b}:&&a_{t10}({1\over \sqrt{3}} \pi^- \Xi^0_{10} - {1\over \sqrt{6}}\pi^0 \Xi^-_{10} + {1\over \sqrt{2}} \eta_8 \Xi^-_{10} + K^0 \Omega^-_{10} - {1\over \sqrt{6}}K^-\Sigma^0_{10} - {1\over \sqrt{3}}\bar K^0 \Sigma^-_{10})\;.\nonumber
\end{eqnarray}

Since there is only one unknown invariant amplitude $a_{t10}$, all $\Delta S = -1$ decays are related among themselves. Also for $\Delta S = 0$ decays. There are also relations between $\Delta S = -1$ and $\Delta S = 0$ decays. For example, indicating the decay width of $\Lambda^0_b \to {\cal P} + {\cal M}$ by $\Gamma^{\Lambda^0_b}_{{\cal P}{\cal M}}$, we have
\begin{eqnarray}
&&\Gamma^{\Lambda^0_b}_{\pi^+ \Delta^-_{10}}:\Gamma^{\Lambda^0_b}_{K^+ \Sigma_{10}^-}:\Gamma^{\Lambda^0_b}_{\pi^0\Delta^0_{10}}:\Gamma^{\Lambda^0_{b}}_{\pi^- \Delta^+_{10}}:\Gamma^{\Lambda^0_b}_{K^0\Sigma^0_{10}}=1:{1\over 3}:{2\over 3}:{1\over 3}:{1\over 6}\;,\nonumber\\
&&\Gamma^{\Lambda^0_b}_{\pi^0\Sigma^0_{10}}:\Gamma^{\Lambda^0_b}_{\pi^+ \Sigma_{10}^-} :\Gamma^{\Lambda^0_{b} }_{K^+ \Xi^-_{10}}: \Gamma^{\Lambda^0_b}_{\pi^- \Sigma^+_{10}}:\Gamma^{\Lambda^0_b}_{K^0\Xi^0_{10}} ={1\over 3}:{1\over 3}:{1\over 3}:{1\over 3}:{1\over 3}\;,
\end{eqnarray}
and
\begin{eqnarray}
{\Gamma^{\Lambda^0_b}_{ K^+ \Sigma_{10}^-}\over \Gamma^{\Lambda^0_{b}}_{\pi^+ \Sigma^-_{10}}} =  {|V_{cd}|^2\over |V_{cs}|^2}\;.
\end{eqnarray}
Similar one can work out relations for $\Xi^{0,-}$ decays.

With more data from  LHCb, these relations can be tested and provide important information for diquark model for pentaquark.

The b-baryon $\Omega^-_b$ is a member in the  sextet ${\cal C}$. Expanding the second equation in eq.(\ref{decuplet}), we obtain, for $\Delta S = 0$ amplitudes,
\begin{eqnarray}
\Omega^-_{b}:&&a_{s10}({1\over\sqrt{6}} K^- \Sigma^0_{10} +{1\over \sqrt{3}} \bar K^0 \Sigma^-_{10} - {\sqrt{2}\over 3}\eta_8 \Xi^0_{10})\nonumber\\
&&+b_{s10}({1\over\sqrt{3}}\pi^- \Xi^0_{10} - {1\over \sqrt{6}}\pi^0\Xi^-_{10} + {1\over 3 \sqrt{2}}\eta_8 \Xi^-_{10} +  K^0 \Omega^-_{10})\;,
\end{eqnarray}
and, for $\Delta S = -1$ amplitudes,
\begin{eqnarray}
\Omega^-_{b}:&&(a_{s10}+b_{s10})({1\over \sqrt{3}} K^- \Xi^0_{10} + {1\over \sqrt{3}} \bar K^0 \Xi^-_{10} - {\sqrt{2}\over \sqrt{3}}\eta_8 \Omega^-_{10})\;.
\end{eqnarray}
Since there are two invariant amplitudes, $a_s$ and $b_s$, there are no simple relations for $\Delta S = 0$ and $\Delta S = -1$ amplitudes, but within each of them, there are direct simple relations and can be tested. We have
\begin{eqnarray}
&&\Gamma^{\Omega^-_b}_{K^-\Sigma^0_{10}}: \Gamma^{\Omega^-_b}_{\bar K^0\Sigma^-_{10}}={1\over 6}:{1\over 3}\;,\nonumber\\
&&\Gamma^{\Omega^-_b}_{\pi^-\Xi^0_{10}}: \Gamma^{\Omega^-_b}_{\pi^0\Xi^-_{10}}:\Gamma^{\Omega^-_b}_{K^0\Omega^-_{10}}={1\over 3}:{1\over 6}:1\;,\nonumber\\
&&\Gamma^{\Omega^-_b}_{K^-\Xi^0_{10}}: \Gamma^{\Omega^-_b}_{\bar K^0\Xi^-_{10}}={1\over 3}:{1\over 3}\;.
\end{eqnarray}

In summary, we have studied some properties of the diquark model for explanations of the $3/2^+$ and $5/2^-$ pentaquarks discovered at the LHCb. In the diquark model, these two pentaquarks are in octet multiplets of flavor $SU(3)$. There is also an additional decuplet pentaquark multiplet and a singlet pentaquark. Finding the states in these multiplets can provide crucial evidence for this model.  The weak decays of b-baryon to a light meson and a pentaquark can have Cabibbo allowed and suppressed decay channels. We find that in the $SU(3)$ limit, for $U$-spin related decay modes the ratio of the decay rates of Cabibbo suppressed to Cabibbo allowed decay channels is given by $|V_{cd}|^2/|V_{cs}|^2$. There are also other testable relations for b-baryon weak decays. These relations can be used as tests for the diquark model for pentaquark.

\begin{acknowledgments}

The work was supported in part by MOE Academic Excellent Program (Grant No: 102R891505) and MOST of ROC, and in part by NSFC(Grant No:11175115) and Shanghai Science and Technology Commission (Grant No: 11DZ2260700) of PRC.

\end{acknowledgments}

\newpage
\appendix

\section{$SU(3)$ amplitudes for a ${\cal B}$ or a ${\cal C}$ decays into an octet pentaquark.}

\begin{table}[!h]
\begin{center}
\caption{$SU(3)$ amplitudes for $\Lambda^{0}_{b}$ decays.}
\begin{tabular}{l|cccccc}
\hline\hline
{Decay mode} & $a_{t}(\overline{3})$ & $b_{t}(\overline{3})$ & $c_{t}(\overline{3})$ &$d_{t}(\overline{3})$
&$e_{t}(\overline{3})$ &  $f_{t}(\overline{3})$ \\
\hline
$\Delta S = 0$&&&&&&\\
\hline
{$\Lambda^{0}_{b} \to K^{+} \Sigma^{-}_{8}$} & (0 &2 &0 &1 &0 & -1)\\
\hline

{$\Lambda^{0}_{b} \to \pi^{-} p_{8}$} & (0 &0 &2 &1 &1  &0 )\\
\hline

{$\Lambda^{0}_{b} \to \eta_{8} n_{8} $} &${1 \over \sqrt{6}}$ (0 &-4 &2 & -1& 1 &2  )\\
\hline

{$\Lambda^{0}_{b} \to K^{0} \Lambda^{0}_{8} $} & ${1 \over \sqrt{6}}$(0 &2 &-4 &-1 &-2  &-1  )\\
\hline

{$\Lambda^{0}_{b} \to K^{0} \Sigma^{0}_{8} $} & ${1 \over \sqrt{2}}$(0 & -2& 0&-1 &0  &1  )\\
\hline

{$\Lambda^{0}_{b} \to \pi^{0} n_{8} $ } &${1 \over \sqrt{2}}$ (0 &0 &-2 &-1 &-1 &  0 )\\
\hline
$\Delta S = -1$&&&&&&\\
\hline

{$\Lambda^{0}_{b} \to K^{+} \Xi^{-}_{8}$} & ( 2 &2 &0  &0 &0 &  -1 )\\ \hline

{$\Lambda^{0}_{b} \to \pi^{+} \Sigma^{-}_{8}$} & (2  & 0& 0 &-1 &0  &0 )\\ \hline

{$\Lambda^{0}_{b} \to K^{-} p_{8}$ } & ( 2 &0 &2  &0 &1 &0  )\\ \hline

{$\Lambda^{0}_{b} \to \pi^{-} \Sigma^{+}_{8}$} & ( 2 &0 &0  &-1 &0  & 0 )\\ \hline

{$\Lambda^{0}_{b} \to K^{0} \Xi^{0}_{8}$ }& (  2 & 2 & 0  &0 &0  &-1  )\\ \hline

{$\Lambda^{0}_{b} \to \bar{K}^{0} n_{8}$} & ( 2 &0 & 2 &0 &1  &0  )\\ \hline

{$\Lambda^{0}_{b} \to \pi^{0} \Sigma^{0}_{8}$} & ( 2 &0 &0  &-1 & 0&0  )\\ \hline

{$\Lambda^{0}_{b} \to \eta_{8} \Lambda^{0}_{8}$} & ${1 \over 3}$( 6 & 4& 4 &1 &2  & -2)\\ \hline

{$\Lambda^{0}_{b} \to \eta_{8} \Sigma^{0}_{8}$} & ( 0 &0 &0  & 0& 0&0  )\\ \hline

{$\Lambda^{0}_{b} \to \pi^{0} \Lambda^{0}_{8}$} & ( 0 &0 &0  &0 &0  &0  )\\
\hline
\hline
\end{tabular}
\label{tab:B3S1}
\end{center}
\end{table}

\begin{table}[!h]
\begin{center}
\caption{$SU(3)$ amplitudes for $\Xi^{0}_{b}$ decays.}
\begin{tabular}{l|cccccc}
\hline\hline
{Decay mode} & $a_{t}(\overline{3})$ & $b_{t}(\overline{3})$ & $c_{t}(\overline{3})$ &$d_{t}(\overline{3})$
&$e_{t}(\overline{3})$ &  $f_{t}(\overline{3})$ \\
\hline
$\Delta S = 0$&&&&&&\\
\hline
{$\Xi^{0}_{b} \to K^{+} \Xi^{-}_{8}$} &( -2 & 0 &0   &1  &0     & 0  )\\
\hline
{$\Xi^{0}_{b} \to \pi^{+} \Sigma^{-}_{8} $ }&( -2 & -2 & 0  &0  &0    &1    )\\
\hline
{$\Xi^{0}_{b} \to K^{-} p_{8}$} &( -2 & 0 & 0  &1  &0   &0   )\\
\hline
{$\Xi^{0}_{b} \to \pi^{-} \Sigma^{+}_{8}$}&(-2  & 0 & -2  &0  &-1   &0   )\\
\hline
{$\Xi^{0}_{b} \to K^{0} \Xi^{0}_{8}$ } &( -2 & 0 & -2  & 0 & -1   & 0  )\\
\hline
{$\Xi^{0}_{b} \to \bar{K}^{0} n_{8}$   } &( -2 & -2 & 0  &0  &0    &1    )\\
\hline
{$\Xi^{0}_{b} \to  \eta_{8} \Lambda^{0}_{8}$ } & ${1 \over 6}$(-12  &-2  &-2   & 4 & -1   & 1 )\\
\hline
{$\Xi^{0}_{b} \to  \eta_{8} \Sigma^{0}_{8}$} &${1 \over 2\sqrt{3}}$ ( 0 &2  &2   &2  & 1   &-1    )\\
\hline
{$\Xi^{0}_{b} \to  \pi^{0} \Lambda^{0}_{8}$ }& ${1 \over 2\sqrt{3}}$   (0  & 2 &  2 & 2 & 1   & -1  )\\
\hline
{$\Xi^{0}_{b} \to  \pi^{0} \Sigma^{0}_{8}$}  & ${1 \over 2}$ ( -4 &-2  &-2   & 0 & -1    &1   )\\
\hline
$\Delta S= -1$&&&&&&\\
\hline

{$\Xi^{0}_{b} \to \pi^{+} \Xi^{-}_{8}$ }& ( 0 &-2 &0  &-1 &0  &1  )\\ \hline

{$\Xi^{0}_{b} \to K^{-} \Sigma^{+}_{8} $} & ( 0 &0 &-2  &-1 &  -1&0  )\\ \hline

{$\Xi^{0}_{b} \to \eta_{8} \Xi^{0}_{8} $} &$ {1\over \sqrt{6}}$ (0  &-2 & 4 &1 &2  &1  )\\ \hline

{$\Xi^{0}_{b} \to \bar{K}^{0} \Lambda^{0}_{8} $} & $ {1\over \sqrt{6}}$ (0  &4 &-2  &1 &-1  & -2)\\ \hline

{$\Xi^{0}_{b} \to \bar{K}^{0} \Sigma^{0}_{8} $}& $ {1\over \sqrt{2}}$( 0 &0 &2  & 1& 1 &0  )\\ \hline

{$\Xi^{0}_{b} \to \pi^{0} \Xi^{0}_{8} $}& $ {1\over \sqrt{2}}$(0  &2 &0  &1 & 0 &-1 )\\
\hline
\hline
\end{tabular}
\label{tab:B2S1}
\end{center}
\end{table}

\begin{table}[!h]
\begin{center}
\caption{$SU(3)$ amplitudes for $\Xi^{-}_{b}$ decays.}
\begin{tabular}{l|cccccc}
\hline\hline
{Decay mode} & $a_{t}(\overline{3})$ & $b_{t}(\overline{3})$ & $c_{t}(\overline{3})$ &$d_{t}(\overline{3})$
&$e_{t}(\overline{3})$ &  $f_{t}(\overline{3})$ \\
\hline
$\Delta S = 0$ &&&&&&\\
\hline
{$\Xi^{-}_{b} \to K^{-} n_{8}$} &( 0 & 2 & 0 & 1 & 0   & -1 )\\
\hline
{$\Xi^{-}_{b} \to K^{0} \Xi^{-}_{8} $ }& ( 0 & 0  & 2 &1  &1    &0 )\\
\hline
{$\Xi^{-}_{b} \to  \eta_{8} \Sigma^{-}_{8}$ }& ${1 \over \sqrt{6}}$ ( 0& 2 & 2 &2  &1    &-1   )\\
\hline
{$\Xi^{-}_{b} \to  \pi^{-} \Lambda^{0}_{8}$ }& ${1 \over \sqrt{6}}$ ( 0& 2 &2  &2  &1    &-1   )\\
\hline
{$\Xi^{-}_{b} \to  \pi^{-} \Sigma^{0}_{8} $} & ${1 \over \sqrt{2}}$ ( 0& -2 & 2 & 0 &1   &1    )\\
\hline
{$\Xi^{-}_{b} \to  \pi^{0} \Sigma^{-}_{8} $ }& ${1 \over \sqrt{2}}$ ( 0& 2 & -2 & 0  & -1   & -1  )\\
\hline
$\Delta S = -1$&&&&&&\\
\hline

{$\Xi^{-}_{b} \to \pi^{-}  \Xi^{0}_{8}$} & (0  &2 & 0  &1 &0  &-1  )\\ \hline

{$\Xi^{-}_{b} \to \bar{K}^{0} \Sigma^{-}_{8} $ } & (0  &0 &2  &1 & 1 &0  )\\ \hline

{$\Xi^{-}_{b} \to  \eta_{8} \Xi^{-}_{8}$} & ${1 \over \sqrt{6}}$ (0  &2 & -4 &-1 &-2  &-1  )\\ \hline

{$\Xi^{-}_{b} \to  K^{-} \Lambda^{0}_{8}$} & ${1 \over \sqrt{6}}$(0  &-4 & 2 &-1 &1 & 2)\\ \hline

{$\Xi^{-}_{b} \to  K^{-} \Sigma^{0}_{8}$} & ${1 \over \sqrt{2}}$(0  &0 & 2 &1 &1  &0 )\\ \hline

{$\Xi^{-}_{b} \to  \pi^{0} \Xi^{-}_{8}$} & ${1 \over \sqrt{2}}$ ( 0 &2 &0  &1 &0  & -1 )\\
\hline
\hline
\end{tabular}
\label{tab:B1S1}
\end{center}
\end{table}

\begin{table}[!h]
\begin{center}
\caption{$SU(3)$ amplitudes for $\Omega^{-}_{b}$ decays.}
\begin{tabular}{l|ccc}
\hline\hline
{Decay mode} & $d_{s}(\overline{3})$ & $e_{s}(\overline{3})$  &$f_{s}(\overline{3})$
 \\
\hline
$\Delta S = 0$&&&\\
\hline
{$\Omega^{-}_{b} \to \eta_{8}\Xi^{-}_{8} $} &${1\over \sqrt{6}}$ (-2 &-1  &0  ) \\ \hline

{$\Omega^{-}_{b} \to K^{-} \Lambda^{0}_{8}$} & ${1\over \sqrt{6}}$(2 &0  &1  ) \\ \hline

{$\Omega^{-}_{b} \to \  \pi^{-} \Xi^{0}_{8}$} & (0 &1  &0  ) \\ \hline

{$\Omega^{-}_{b} \to \  \pi^{0} \Xi^{-}_{8}$} &${1\over \sqrt{2}}$ ( 0&1  &0  ) \\ \hline

{$\Omega^{-}_{b} \to K^{-} \Sigma^{0}_{8}$} & ${1\over\sqrt{2}}$(0 &0  &-1  ) \\ \hline

{$\Omega^{-}_{b} \to \Sigma^{-}_8 \bar{K}^{0}$} & (0 &0  &-1  ) \\

\hline
$\Delta S = -1$&&&\\
\hline
{$\Omega^{-}_{b} \to  K^{-} \Xi^{0}_{8}$} & (1 &1  &1  ) \\ \hline

{$\Omega^{-}_{b} \to  \bar{K}^{0} \Xi^{-}_{8}$} & ( -1& -1 &-1  ) \\
\hline
\hline
\end{tabular}
\label{tab:B33S1}
\end{center}
\end{table}

\begin{table}[!h]
\begin{center}
\caption{$SU(3)$ amplitudes for $\Xi'^{0}_{b}$ decays.}
\begin{tabular}{l|ccc}
\hline\hline
{Decay mode} & $d_{s}(\overline{3})$ & $e_{s}(\overline{3})$  &$f_{s}(\overline{3})$
 \\
\hline
$\Delta S = 0$&&&\\
\hline
{$ \Xi'^{0}_{b} \to K^{+} \Xi^{-}_{8} $} &  ${1\over\sqrt{2}}$(1 &0  &0  ) \\ \hline

{$ \Xi'^{0}_{b} \to \pi^{+} \Sigma^{-}_{8} $} & ${1\over\sqrt{2}}$ (0 & 0& -1) \\ \hline

{$ \Xi'^{0}_{b} \to K^{-} p_{8} $} & ${1\over\sqrt{2}}$ (-1 &0  &0  ) \\ \hline

{$ \Xi'^{0}_{b} \to \pi^{-} \Sigma^{+}_{8} $} &  ${1\over\sqrt{2}}$( 0&1  &0  ) \\ \hline

{$ \Xi'^{0}_{b} \to \eta_{8} \Lambda^{0}_{8} $} &  ${1\over2\sqrt{2}}$(0 &-1  &1  ) \\ \hline

{$ \Xi'^{0}_{b} \to K^{0} \Xi^{0}_{8} $} & ${1\over\sqrt{2}}$ (0 &-1  &0  ) \\ \hline

{$ \Xi'^{0}_{b} \to \eta_{8} \Sigma^{0}_{8} $} &  ${1\over 2\sqrt{6}}$( -2& -1 &-3  ) \\ \hline

{$ \Xi'^{0}_{b} \to \bar{K}^{0} n_{8} $} &  ${1\over \sqrt{2}}$(0 &0  &1  ) \\ \hline

{$ \Xi'^{0}_{b} \to \pi^{0} \Lambda^{0}_{8} $} &  ${1\over 2\sqrt{6}}$( 2&3  &1  ) \\ \hline

{$ \Xi'^{0}_{b} \to \pi^{0} \Sigma^{0}_{8} $} & ${1\over 2\sqrt{2}}$ (0 &1  &-1  ) \\
\hline
$\Delta S = -1$&&&\\
\hline
{$\Xi'^{0}_{b} \to  \pi^{+} \Xi^{-}_{8}$} &  ${1\over\sqrt{2}}$(-1 &0  &-1  ) \\ \hline

{$\Xi'^{0}_{b} \to  K^{-} \Sigma^{+}_{8}$} &  ${1\over\sqrt{2}}$( 1&1  &0  ) \\ \hline

{$\Xi'^{0}_{b} \to \eta_{8} \Xi^{0}_{8}$} & ${1\over 2\sqrt{3}}$ (1 &2  & 3 ) \\ \hline

{$\Xi'^{0}_{b} \to \bar{K}^{0} \Lambda^{0}_{8}$} & ${1\over 2\sqrt{3}}$ ( -1& -3&-2  ) \\ \hline

{$\Xi'^{0}_{b} \to \bar{K}^{0} \Sigma^{0}_{8}$} &  ${1\over 2}$(-1 &-1  &0  ) \\ \hline

{$\Xi'^{0}_{b} \to \pi^{0} \Xi^{0}_{8}$} & ${1\over 2 }$ (1 & 0&1 ) \\
\hline
\hline
\end{tabular}
\label{tab:B13S1}
\end{center}
\end{table}

\begin{table}[!h]
\begin{center}
\caption{$SU(3)$ amplitudes for $\Xi'^{-}_{b}$ decays.}
\begin{tabular}{l|ccc}
\hline\hline
{Decay mode} & $d_{s}(\overline{3})$ & $e_{s}(\overline{3})$  &$f_{s}(\overline{3})$
 \\
\hline
$\Delta S = 0$&&&\\
\hline
{$ \Xi'^{-}_{b} \to \eta_{8} \Sigma^{-}_{8} $} &  ${1\over 2\sqrt{3}}$( -2& -1 &-3  ) \\ \hline

{$ \Xi'^{-}_{b} \to K^{-} n_{8} $} &  ${1\over\sqrt{2}}$ (-1 &0  &-1  ) \\ \hline

{$ \Xi'^{-}_{b} \to K^{0} \Xi^{-}_{8} $} &  ${1\over\sqrt{2}}$(1 &1  &0  ) \\ \hline

{$ \Xi'^{-}_{b} \to \pi^{-}\Lambda^{0}_{8} $}& ${1\over 2\sqrt{3}}$ ( 2&3  &1  ) \\ \hline

{$ \Xi'^{-}_{b} \to \pi^{-}\Sigma^{0}_{8} $} & ${1\over 2}$  (0 &-1  &-1  ) \\ \hline

{$ \Xi'^{-}_{b} \to \pi^{0}\Sigma^{-}_{8} $}&  ${1\over 2}$(0 &1  &1  ) \\
\hline
$\Delta S = -1$&&&\\
\hline
{$\Xi'^{-}_{b} \to  \eta_{8} \Xi^{-}_{8}$} &  ${1 \over 2\sqrt{3}}$(-1 &-2  &-3  ) \\ \hline

{$\Xi'^{-}_{b} \to  K^{-} \Lambda^{0}_{8}$} &  ${1\over 2\sqrt{3}}$( 1& 3& 2 ) \\ \hline

{$\Xi'^{-}_{b} \to \pi^{-} \Xi^{0}_{8}$} &  ${1\over\sqrt{2}}$( 1& 0&1  ) \\ \hline

{$\Xi'^{-}_{b} \to K^{-} \Sigma^{0}_{8}$} &  ${1\over 2}$( -1& -1& 0) \\ \hline

{$\Xi'^{-}_{b} \to \bar{K}^{0} \Sigma^{-}_{8}$} &  ${1\over\sqrt{2}}$(-1 &-1  &0  ) \\ \hline

{$\Xi'^{-}_{b} \to \pi^{0} \Xi^{-}_{8}$} &  ${1\over 2}$( 1&0  &1 ) \\
\hline
\hline
\end{tabular}
\label{tab:B23S1}
\end{center}
\end{table}

\begin{table}[!h]
\begin{center}
\caption{$SU(3)$ amplitudes for $\Sigma^{+}_{b}$ decays.}
\begin{tabular}{l|ccc}
\hline\hline
{Decay mode} & $d_{s}(\overline{3})$ & $e_{s}(\overline{3})$  &$f_{s}(\overline{3})$
 \\
\hline
$\Delta S = 0$&&&\\
\hline
{$\Sigma^{+}_{b} \to  \eta_{8} p_{8}$} &${1\over \sqrt{6}}$ (-1 &1  &0 )\\ \hline

{$\Sigma^{+}_{b} \to \pi^{+} n_{8}$}  & (0 &0  &1  )\\ \hline

{$\Sigma^{+}_{b} \to K^{0} \Sigma^{+}_{8}$}  & (0 &-1  &0  )\\ \hline

{$\Sigma^{+}_{b} \to K^{+} \Lambda^{0}_{8}$}  & ${1\over \sqrt{6}}$ ( 1&0  &-1  )\\ \hline

{$\Sigma^{+}_{b} \to K^{+} \Sigma^{0}_{8}$}  &${1\over \sqrt{2}}$  (1 &0 & 1 )\\ \hline

{$\Sigma^{+}_{b} \to p_8 \pi^{0}$}  & ${1\over \sqrt{2}}$ ( -1&-1  &0  )\\
\hline
$\Delta S = -1$&&&\\
\hline
{$\Sigma^{+}_{b} \to  \eta_{8} \Sigma^{+}_{8}$} & ${1\over\sqrt{6}}$ (1 &2  &0  ) \\ \hline

{$\Sigma^{+}_{b} \to  \pi^{+} \Lambda^{0}_{8}$} &  ${1\over\sqrt{6}}$( -1& 0&-2  ) \\ \hline

{$\Sigma^{+}_{b} \to  K^{+} \Xi^{0}_{8}$} & (0 &0  &-1  ) \\ \hline

{$\Sigma^{+}_{b} \to  \pi^{+} \Sigma^{0}_{8}$} &  ${1\over\sqrt{2}}$( -1&0  &0  ) \\ \hline

{$\Sigma^{+}_{b} \to  \bar{K}^{0} p_{8}$} & (0 &1  &0  ) \\ \hline

{$\Sigma^{+}_{b} \to  \pi^{0} \Sigma^{+}_{8}$} &  ${1\over\sqrt{2}}$( 1& 0 &0  ) \\
\hline
\hline
\end{tabular}
\label{tab:B11S1}
\end{center}
\end{table}

\begin{table}[!h]
\begin{center}
\caption{$SU(3)$ amplitudes for $\Sigma^{0}_{b}$ decays.}
\begin{tabular}{l|ccc}
\hline\hline
{Decay mode} & $d_{s}(\overline{3})$ & $e_{s}(\overline{3})$  &$f_{s}(\overline{3})$
 \\
\hline
$\Delta S = 0$&&&\\
\hline
{$\Sigma^{0}_{b} \to \eta_{8} n_{8}$}&  ${1\over 2\sqrt{3}}$ (-1 &1  &0  ) \\ \hline

{$\Sigma^{0}_{b} \to K^{+} \Sigma^{-}_{8}$} &  ${1\over\sqrt{2}}$( 1& 0&1  ) \\ \hline

{$\Sigma^{0}_{b} \to \pi^{-} p_{8} $} & ${1\over\sqrt{2}}$ ( -1&-1 &0  ) \\ \hline

{$\Sigma^{0}_{b} \to K^{0} \Lambda^{0}_{8} $} & ${1\over 2\sqrt{3}}$ ( 1&0  &-1  ) \\ \hline

{$\Sigma^{0}_{b} \to K^{0} \Sigma^{0}_{8} $} & ${1 \over 2}$ (1 &2  &1  ) \\ \hline

{$\Sigma^{0}_{b} \to n_{8} \pi^{0} $}& ${1\over 2}$(-1 & -1& -2 ) \\
\hline
$\Delta S = -1$&&&\\
\hline
{$\Sigma^{0}_{b} \to  K^{+} \Xi^{-}_{8}$} & ${1\over\sqrt{2}}$ (0 &0  &1  ) \\ \hline

{$\Sigma^{0}_{b} \to  \pi^{+} \Sigma^{-}_{8}$} & ${1\over\sqrt{2}}$ ( -1& 0& 0 ) \\ \hline

{$\Sigma^{0}_{b} \to  K^{-} p_{8}$} &  ${1\over\sqrt{2}}$( 0&-1  &0 ) \\ \hline

{$\Sigma^{0}_{b} \to  \pi^{-} \Sigma^{+}_{8}$} & ${1\over\sqrt{2}}$ (1 &0  &0  ) \\ \hline

{$\Sigma^{0}_{b} \to  \eta_{8} \Lambda^{0}_{8}$} & ( 0&0  &0  ) \\ \hline

{$\Sigma^{0}_{b} \to  K^{0} \Xi^{0}_{8}$} &  ${1\over\sqrt{2}}$(0 &0  &-1  ) \\ \hline

{$\Sigma^{0}_{b} \to  \eta_{8} \Sigma^{0}_{8}$} & ${1\over\sqrt{6}}$ ( -1& -2&0 ) \\ \hline

{$\Sigma^{0}_{b} \to  \bar{K}^{0} n_{8} $} & ${1\over\sqrt{2}}$ ( 0&1  &0  ) \\ \hline

{$\Sigma^{0}_{b} \to  \pi^{0} \Lambda^{0}_{8}$} &  ${1\over\sqrt{6}}$(1 & 0& 2 ) \\ \hline

{$\Sigma^{0}_{b} \to  \pi^{0} \Sigma^{0}_{8}$} & ( 0&0  &0  ) \\
\hline
\hline
\end{tabular}
\label{tab:B12S1}
\end{center}
\end{table}

\begin{table}[!h]
\begin{center}
\caption{$SU(3)$ amplitudes for $\Sigma^{-}_{b} $ decays.}
\begin{tabular}{l|ccc}
\hline\hline
{Decay mode} & $d_{s}(\overline{3})$ & $e_{s}(\overline{3})$  &$f_{s}(\overline{3})$
 \\
\hline
$\Delta S=0$&&&\\
\hline
{$\Sigma^{-}_{b} \to  \pi^{-} n_{8}$} & ( -1& -1 &-1  ) \\ \hline

{$\Sigma^{-}_{b} \to K^{0} \Sigma^{-}_{8}$} & (1 & 1 &1  ) \\

\hline
$\Delta S = -1$&&&\\
\hline
{$\Sigma^{-}_{b} \to  \eta_{8} \Sigma^{-}_{8}$} &  ${1\over\sqrt{6}}$( -1& -2&0  ) \\ \hline

{$\Sigma^{-}_{b} \to  K^{-} n_{8}$}& (0 &-1  &0 ) \\ \hline

{$\Sigma^{-}_{b} \to  K^{0} \Xi^{-}_{8}$}& ( 0& 0 &1  ) \\ \hline

{$\Sigma^{-}_{b} \to \pi^{-} \Lambda^{0}_{8}$}&  ${1\over\sqrt{6}}$( 1&0  &2  ) \\ \hline

{$\Sigma^{-}_{b} \to \pi^{-} \Sigma^{0}_{8}$}&  ${1\over\sqrt{2}}$(-1 &0  &0  ) \\ \hline

{$\Sigma^{-}_{b} \to \pi^{0} \Sigma^{-}_{8}$}& ${1\over\sqrt{2}}$ ( 1& 0 & 0 ) \\
\hline
\hline
\end{tabular}
\label{tab:B22S1}
\end{center}
\end{table}

\section{$SU(3)$ amplitudes for an sextet decays into a decuplet pentaquark}

For $\Delta S = 0$ amplitudes, we have
\begin{eqnarray}
\Sigma^+_b:&&a_{s10}({1\over \sqrt{6}}\pi^0\Delta^+_{10} + {1\over 3\sqrt{2}}\eta_8 \Delta^+_{10} + {1\over \sqrt{3}} \pi^+\Delta^0_{10} + {1\over \sqrt{6}} K^+ \Sigma^0_{10})\nonumber\\
&& +b_{s10} (\pi^-\Delta^{++}_{10} - {1\over \sqrt{6}} \pi^0\Delta^+_{10} +{1\over 3\sqrt{2}}\eta_8 \Delta^+_{10} + {1\over \sqrt{3}} K^0 \Sigma^+_{10})\;,\nonumber\\
\Sigma^0_b:&& a_{s10}({1\over 3}\eta_8 \Delta^0_{10}+ {1\over \sqrt{2}}\pi^+\Delta^-_{10} +{1\over \sqrt{6}}K^+\Sigma^-_{10}+{1\over \sqrt{6}}\pi^- \Delta^+_{10} + {1\over 2\sqrt{3}}K^0 \Sigma^0_{10})\nonumber\\
&& +b_{s10}({\sqrt{2}\over \sqrt{3}} \pi^-\Delta^+_{10} - {1\over \sqrt{3}}\pi^0 \Delta^0_{10} + {1 \over 3}\eta_8 \Delta^0_{10} + {1\over \sqrt{3}} K^0 \Sigma^0_{10})\;,
\nonumber\\
\Sigma^-_{b}: &&(a_{s10}+b_{s10})({1\over \sqrt{3}}\pi^-\Delta^0_{10} -{1\over \sqrt{2} }\pi^0 \Delta^-_{10} + {1\over\sqrt{6}}\eta_8 \Delta^-_{10} +{1\over\sqrt{3}}K^0\Sigma^-_{10})\;,\nonumber\\
\Xi^{'0}_{b}:&& a_{s10}({1\over \sqrt{6}} K^- \Delta^+_{10} + {1\over \sqrt{6}} \bar K^0 \Delta^0_{10}-{1\over 6\sqrt{2}} \eta_8 \Sigma^0_{10} + {1\over 2 \sqrt{6}}\pi^0 \Sigma^0_{10} + {1\over \sqrt{6}} \pi^+ \Sigma^-_{10} +{1\over \sqrt{6}}K^+ \Xi^-_{10})\nonumber\\
&& +b_{s10}({\sqrt{2}\over \sqrt{3}} \pi^- \Sigma^+_{10} - {1\over \sqrt{6}} \pi^0 \Sigma^0_{10} + {\over 3\sqrt{2}}\eta_8 \Sigma^0_{10}+{\sqrt{2} \over \sqrt{3}}K^0 \Xi^0_{10})\;,\nonumber\\
\Xi^{'-}_b:&& a_{s10}({1\over \sqrt{6}}K^- \Delta^0_{10} + {1 \over \sqrt{2}}\bar K^0 \Delta^-_{10} +{1\over 2\sqrt{3}}\pi^- \Sigma^0_{10} - {1\over 2\sqrt{3}} \pi^0 \Sigma^-_{10} - {1\over 6} \eta_8 \Sigma^-_{10} + {1\over \sqrt{6}}K^0 \Xi^-_{10})\nonumber\\
&&+ b_{s10}({1\over \sqrt{3}}\pi^- \Sigma^0_{10} - {1\over \sqrt{3}}\pi^0 \Sigma^-_{10} + {1\over 3}\eta_8 \Sigma^-_{10} + {\sqrt{2}\over \sqrt{3}}K^0 \Xi^-_{10})\nonumber\\
\Omega^-_{b}:&&a_{s10}({1\over\sqrt{6}} K^- \Sigma^0_{10} +{1\over \sqrt{3}} \bar K^0 \Sigma^-_{10} - {\sqrt{2}\over 3}\eta_8 \Xi^0_{10})\nonumber\\
&&+b_{s10}({1\over\sqrt{3}}\pi^- \Xi^0_{10} - {1\over \sqrt{6}}\pi^0\Xi^-_{10} + {1\over 3 \sqrt{2}}\eta_8 \Xi^-_{10} +  K^0 \Omega^-_{10})\;,
\end{eqnarray}
and for $\Delta S = -1$ amplitudes, we have
\begin{eqnarray}
\Sigma^+_b:&&a_{s10}({1\over \sqrt{6}}\pi^0 \Sigma^+_{10} +{1\over 3\sqrt{2}}\eta_8 \Sigma^+_{10} + {1\over \sqrt{6}} \pi^+ \Sigma^0_{10} + {1\over \sqrt{3}}K^+ \Xi^0_{10})\nonumber\\
&&+ b_{s10} (K^- \Delta^{++}_{10} + {1\over \sqrt{3}} \bar K^0 \Delta^+_{10} - {\sqrt{2}\over 3} \eta_8 \Sigma^+_{10})\;,\nonumber\\
\Sigma^0_b:&& a_{s10}({1\over \sqrt{6}}\pi^- \Sigma^+_{10} + {1\over 3\sqrt{2}}\eta_8 \Sigma^0_{10} +{1\over \sqrt{6}}K^0 \Xi^0_{10} +{1\over \sqrt{6}} \pi^+\Sigma^-_{10} + {1\over \sqrt{6}}K^+ \Xi^-_{10})\nonumber\\
&&+  b_{s10}({\sqrt{2}\over\sqrt{3}} K^-\Delta^+_{10} + {\sqrt{2}\over \sqrt{3}} \bar K^0 \Delta^0_{10} -{\sqrt{2}\over 3}\eta_8 \Sigma^0_{10})\;,\nonumber\\
\Sigma^-_{b}:&&a_{s10}({1\over \sqrt{6}} \pi^- \Sigma^0_{10} - {1\over \sqrt{6}} \pi^0 \Sigma^-_{10} + {1\over 3 \sqrt{2}}\eta_8 \Sigma^-_{10} + {1\over \sqrt{3}}  K^0 \Xi^-_{10})\nonumber\\
&&+b_{s10}({1\over \sqrt{3}} K^- \Delta^0_{10} + \bar K^0 \Delta^-_{10}
- {\sqrt{2}\over 3}\eta_8 \Sigma^-_{10})\;,\nonumber\\
\Xi^{'0}_b:&& a_{s10} ({1\over \sqrt{6}} K^- \Sigma^+_{10} + {1\over 2\sqrt{3}} \bar K^0 \Sigma^0_{10} +{1\over 2\sqrt{3}}\pi^0 \Xi^0_{10} - {1\over 6} \eta_8 \Xi^0_{10} + {1\over \sqrt{6}} \pi^+ \Xi^-_{10} + {1\over \sqrt{2}}K^+ \Omega^-_{10})\nonumber\\
&& b_{s10} ({\sqrt{2}\over \sqrt{3}}K^-\Sigma^+_{10} + {1\over\sqrt{3}}\bar K^0 \Sigma^0_{10} - {2\over 3} \eta_8 \Xi^0_{10})\;,\nonumber\\
\Xi^{'-}_b:&& a_{s10} ({1\over 2\sqrt{3}}K^-\Sigma^0_{10} +{1\over \sqrt{6}}\bar K^0 \Sigma^-_{10} -{1\over 6}\eta_8 \Xi^-_{10} +{1\over \sqrt{6}}\pi^- \Xi^0_{10} - {1\over 2\sqrt{3}}\pi^0 \Xi^-_{10} + {1\over \sqrt{2}}K^0 \Omega^-_{10})\nonumber\\
&&+ b_{s10} ({1\over \sqrt{3}}K^- \Sigma^0_{10} + {\sqrt{2}\over \sqrt{3}} \bar K^0\Sigma^-_{10} - {2\over 3} \eta_8 \Xi^-_{10})\;, \nonumber\\
\Omega^-_{b}:&&(a_{s10}+b_{s10})({1\over \sqrt{3}} K^- \Xi^0_{10}+ {1\over \sqrt{3}} \bar K^0 \Xi^-_{10} - {\sqrt{2}\over \sqrt{3}}\eta_8 \Omega^-_{10})\;.
\end{eqnarray}


\begin{thebibliography}{99}

\bibitem{Aaij:2015tga}
  R.~Aaij {\it et al.} [LHCb Collaboration],
  arXiv:1507.03414 [hep-ex].

\bibitem{rosner}
M.~Karliner and J.~L.~Rosner,
  Phys.\ Rev.\ Lett.\  {\bf 115}, no. 12, 122001 (2015)
  [arXiv:1506.06386 [hep-ph]].

\bibitem{Chen:2015loa}
  R.~Chen, X.~Liu, X.~Q.~Li and S.~L.~Zhu,
  arXiv:1507.03704 [hep-ph].

\bibitem{Chen:2015moa}
  H.~X.~Chen, W.~Chen, X.~Liu, T.~G.~Steele and S.~L.~Zhu,
  arXiv:1507.03717 [hep-ph].

\bibitem{Roca:2015dva}
  L.~Roca, J.~Nieves and E.~Oset,
  arXiv:1507.04249 [hep-ph].

\bibitem{Feijoo:2015cca}
  A.~Feijoo, V.~K.~Magas, A.~Ramos and E.~Oset,
  arXiv:1507.04640 [hep-ph].

\bibitem{Mironov:2015ica}
  A.~Mironov and A.~Morozov,
  arXiv:1507.04694 [hep-ph].

\bibitem{Guo:2015umn}
  F.~K.~Guo, U.~G.~Meißner, W.~Wang and Z.~Yang,
  arXiv:1507.04950 [hep-ph].

\bibitem{Maiani:2015vwa}
  L.~Maiani, A.~D.~Polosa and V.~Riquer,
  arXiv:1507.04980 [hep-ph].

\bibitem{He:2015cea}
  J.~He,
  arXiv:1507.05200 [hep-ph].

\bibitem{Liu:2015fea}
  X.~H.~Liu, Q.~Wang and Q.~Zhao,
  arXiv:1507.05359 [hep-ph].


\bibitem{Lebed:2015tna}
  R.~F.~Lebed,
  arXiv:1507.05867 [hep-ph].

\bibitem{Mikhasenko:2015vca}
  M.~Mikhasenko,
  arXiv:1507.06552 [hep-ph].

\bibitem{xc}
Ulf-G. Meisner, Jose A. Oller, arXiv:1507.07525.

\bibitem{diquark1}
V. Anisovich, M. Matveev, J. Nyiri, A. Sarantsev and A. Semenova, arXiv:1507.07652.

\bibitem{heff}
G.~Buchalla, A.~J.~Buras and M.~E.~Lautenbacher,
  Rev.\ Mod.\ Phys.\  {\bf 68}, 1125 (1996).

\bibitem{b-baryon}
W.~Roberts and M.~Pervin,
  Int.\ J.\ Mod.\ Phys.\ A {\bf 23}, 2817 (2008).

\bibitem{he-b-baryon}
  X.~G.~He and G.~N.~Li,
  Phys.\ Lett.\ B {\bf 750}, 82 (2015)
  [arXiv:1501.00646 [hep-ph]];
M.~He, X.~G.~He and G.~N.~Li,
  Phys.\ Rev.\ D {\bf 92}, no. 3, 036010 (2015)
  [arXiv:1507.07990 [hep-ph]].


\end{thebibliography}
\end{document}